\documentclass[preprint, preprintnumbers,amsmath,amssymb,superscriptaddress,longbibliography]{revtex4-1}
\usepackage{graphicx}
\usepackage{dcolumn}
\usepackage{bm}
\usepackage{subfigure}
\usepackage{multirow}
\usepackage{amsmath}
\usepackage{color}
\usepackage[font=footnotesize, justification=raggedright,singlelinecheck=false]{caption}
\begin{document}

\title{Melting and re-entrant melting of polydisperse hard disks}

\author{Pablo Sampedro Ruiz}
\affiliation{School of Chemical and Biomedical Engineering, Nanyang Technological University, \\62 Nanyang Drive, 637459, Singapore 
}

\author{Qun-li Lei}
\affiliation{School of Chemical and Biomedical Engineering, Nanyang Technological University, \\62 Nanyang Drive, 637459, Singapore 
}

\author{Ran Ni}
\email{r.ni@ntu.edu.sg}
\affiliation{School of Chemical and Biomedical Engineering, Nanyang Technological University, \\62 Nanyang Drive, 637459, Singapore 
}

\begin{abstract}
Because of long-wavelength fluctuations, the nature of solids and phase transitions in 2D are different from those in 3D systems, and have been heavily debated in past decades, in which the focus was on the existence of hexatic phase. Here, by using large scale computer simulations, we investigate the melting transition in 2D systems of polydisperse hard disks. We find that, with increasing the particle size polydispersity, the melting transition can be qualitatively changed from the recently proposed two-stage process to the Kosterlitz-Thouless-Halperin-Nelson-Young scenario with significantly enlarged stability range for hexatic phase. Moreover, re-entrant melting transitions are found in high density systems of polydisperse hard disks, which were proven impossible in 3D polydisperse hard-sphere systems. These suggest a new fundamental difference between phase transitions in polydisperse systems in 2D and 3D.

\end{abstract}

\maketitle

\section{Introduction}
Melting of two-dimensional solids has been heavily discussed since it was proven that long-wavelength thermal fluctuations prevent the long-range positional order in 2D systems~\cite{Strandburg1988,Dash1999,Grasser2009}.
A significant attempt to reach the general description of 2D melting was the Kosterlitz-Thouless-Halperin-Nelson-Young (KTHNY) theory, that predicts a new phase of matter, i.e., the hexatic phase, which has quasi-long-range orientational order and short-range positional order. The hexatic phase is predicted to be interposed between the usual solid and fluid phases. Therefore, the melting in 2D was predicted to undergo two continuous transitions from solid to hexatic and hexatic to fluid, respectively~\cite{kt,hn2,hn,young,bkt}. However, the KTHNY theory does not rule out the first-order transition due to other effects~\cite{binder2002}, e.g., the  grain-boundary induced melting~\cite{chiu1982,saito1982}. 
In arguably the simplest benchmarking model system in 2D, i.e., the system of monodisperse hard disks, the physics of melting transition has been debated for long~\cite{zahn1999,karn2000,han2008,RICE20091,murray1987,marcus1996,maret2004,keim2007,stuart2008}. It was recently settled that the melting of solids in the system of monodisperse hard disk undergoes a two-stage process consisting of a continuous solid-hexatic transition followed by a first-order hexatic-fluid transition~\cite{hdprl,hdpre}, and the shape and softness of particles also play important roles in the 2D melting~\cite{krauth2015,glotzer2017,massimo2018}. 

It was found that pinning a small fraction of particles can change the melting transition in hard-disk systems significantly~\cite{lowen2013,weikai2015}.
Moreover, simulations of binary hard-disk mixtures showed that the presence of tiny amounts of small particles can eliminate the stability of hexatic phase~\cite{russo2017}. These highlight that the melting transition in 2D is subtle. A recent experiment with colloidal hard spheres in 2D~\cite{dullens2017} appears to support the two-stage melting found in simulation~\cite{hdprl,hdpre}. However, most of experimental systems possess certain degree of (continuous) particle size polydispersity, and polydisperse hard disks have also been widely employed as a model system to investigate the glass transition~\cite{tanaka2007,tanaka2011,tanaka2015}. Yet the effect of polydispersity on the nature of phase transitions in 2D remains unknown.
To this end, we investigate the equilibrium phase behaviour of a 2D polydisperse hard-disk system (PHDS) with Gaussian-like particle size polydispersity. We find that with increasing the polydispersity, the first-order hexatic-fluid transition becomes weaker, and completely switches to be continuous following the KTHNY scenario in PHDS with around $7\%$ size polydispersity. Concurrently, the packing fraction range for stable hexatic phase increases significantly by one order of magnitude compared to that in monodisperse hard-disk systems. More surprisingly, in PHDS with slightly higher polydispersity, we observe re-entrant solid-hexatic and hexatic-fluid transitions at high density, which were proven impossible in 3D polydisperse hard-sphere systems~\cite{sollich2003prl}.

\begin{figure*}
\centering \includegraphics[width=\textwidth]{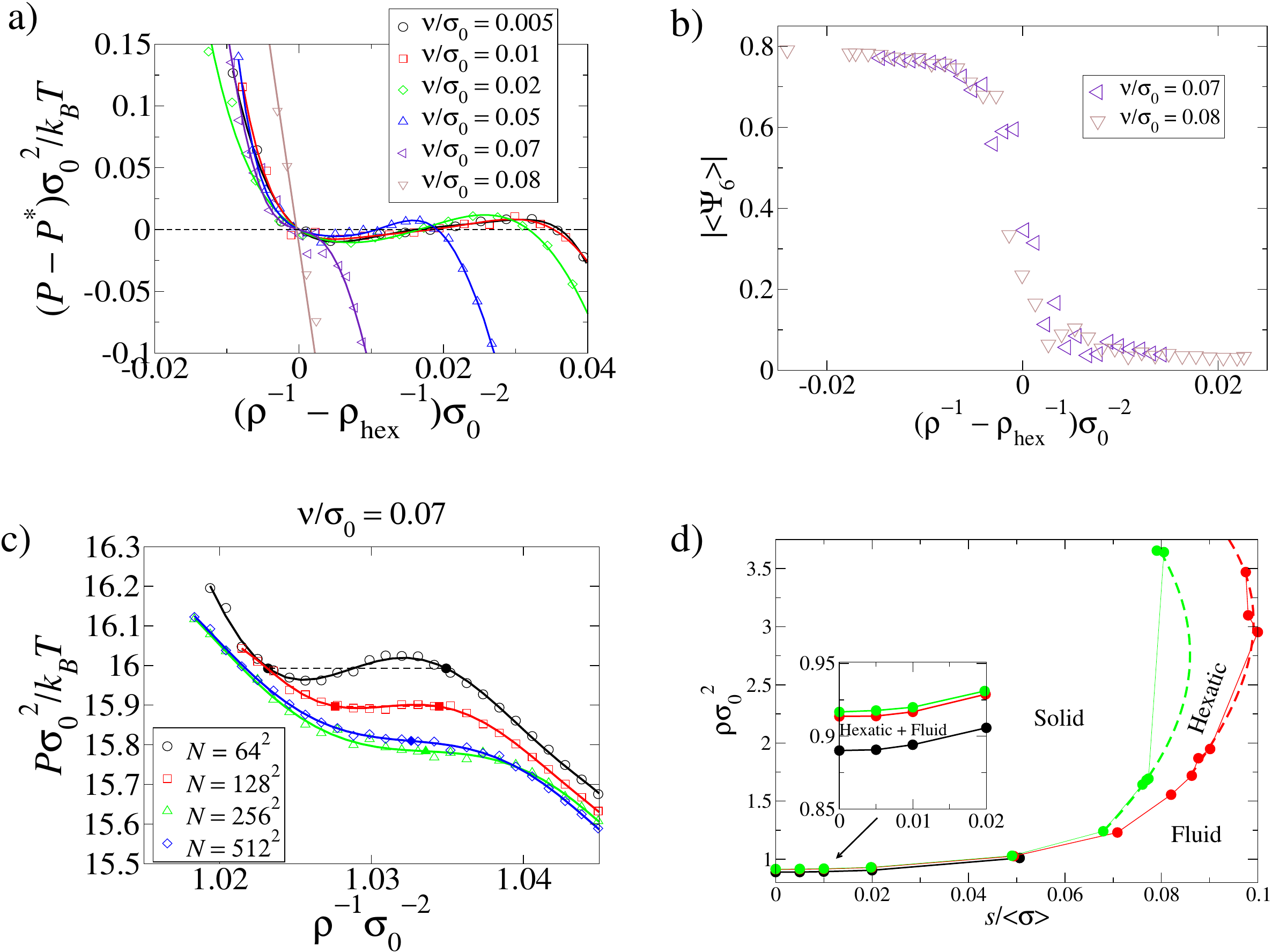}
\caption{{\textbf{Phase behaviour of polydisperse hard disks.}}(a): Equation of state (EOS) for polydisperse hard disk systems (PHDS) with various polydispersity parameter $\nu/\sigma_0 = 0.005$ to $0.08$ in the representation of $(P-P^*)\sigma_0^2/ k_B T$ vs $(\rho^{-1} - \rho_{hex}^{-1})\sigma_0^{-2}$, where $P^*$ and $\rho_{hex}$ are the pressure and density of the hexatic phase at the fluid-hexatic transition, respectively, and the solid lines are fits of the EOS using 5th order polynomials. (b): $\left | \langle \Psi_6 \rangle \right |$ as functions of $(\rho^{-1} - \rho_{hex}^{-1})\sigma_0^{-2}$ for systems with $\nu/\sigma_0 = 0.07$ and 0.08. $P$ and $\rho=N/V$ are the pressure of the system and the density of particles in the system, respectively.
(c): EOSs for PHDS of various numbers of particles $N = 64^2 \sim 512^2$ at $\nu/\sigma_0 = 0.07$. The solid symbols are the obtained fluid-hexatic transition points.
(d): Phase diagram of the PHDS in the representation of $\rho \sigma_0^2$ and $s/\langle \sigma \rangle$, in which the dashed lines are the interpolated phase boundaries for re-entrant melting of solid and hexatic phases. The phase boundaries are obtained from $NVT-\Delta \mu$ MC simulations for PHDS with $\nu/\sigma_0 = 0.005$ to 0.0835. Inset: the enlarged view of the region of phase diagram at $0 \le s / \langle \sigma \rangle \le 0.02$. The dotted lines connect the re-entrant transitions at $\nu/\sigma_0 = 0.08,0.0805,0.081$, $0.082, 0.083$, and 0.0835 from left to right.}
\label{fig1}
\end{figure*} 

\section{Results}
\subsection{Model}
To model the effect of polydispersity, we consider a 2D system of volume $V$ containing $N$ polydisperse hard disks based on the semigrand canonical ensemble, in which the chemical potential difference between particles of different size is fixed~\cite{kofke1988,bolhuis1996,kofke1999,frenkel2004}. In this work, we use the following function for the distribution of chemical potential difference 
\begin{equation}\label{eq1}
\Delta \mu (\sigma) = -k_B T \frac{(\sigma - \sigma_0)^2}{2 \nu^2},
\end{equation}
where $\sigma$ is the particle diameter changing from 0 to $\infty$ with $k_B$ and $T$ the Boltzmann constant and temperature of the system, respectively. $\nu$ is the polydispersity parameter, and in the ideal gas limit Eq.~(\ref{eq1}) gives a Gaussian-like particle size distribution centered around $\sigma_0$ with the standard deviation $\nu$. This models PHDS in equilibrium with the dilute reservoir of Gaussian-like particle size distribution, e.g., the very top region in sedimentation experiments~\cite{dullens2017}. 

\begin{figure}[tb]
\centering \includegraphics[width=0.4\textwidth]{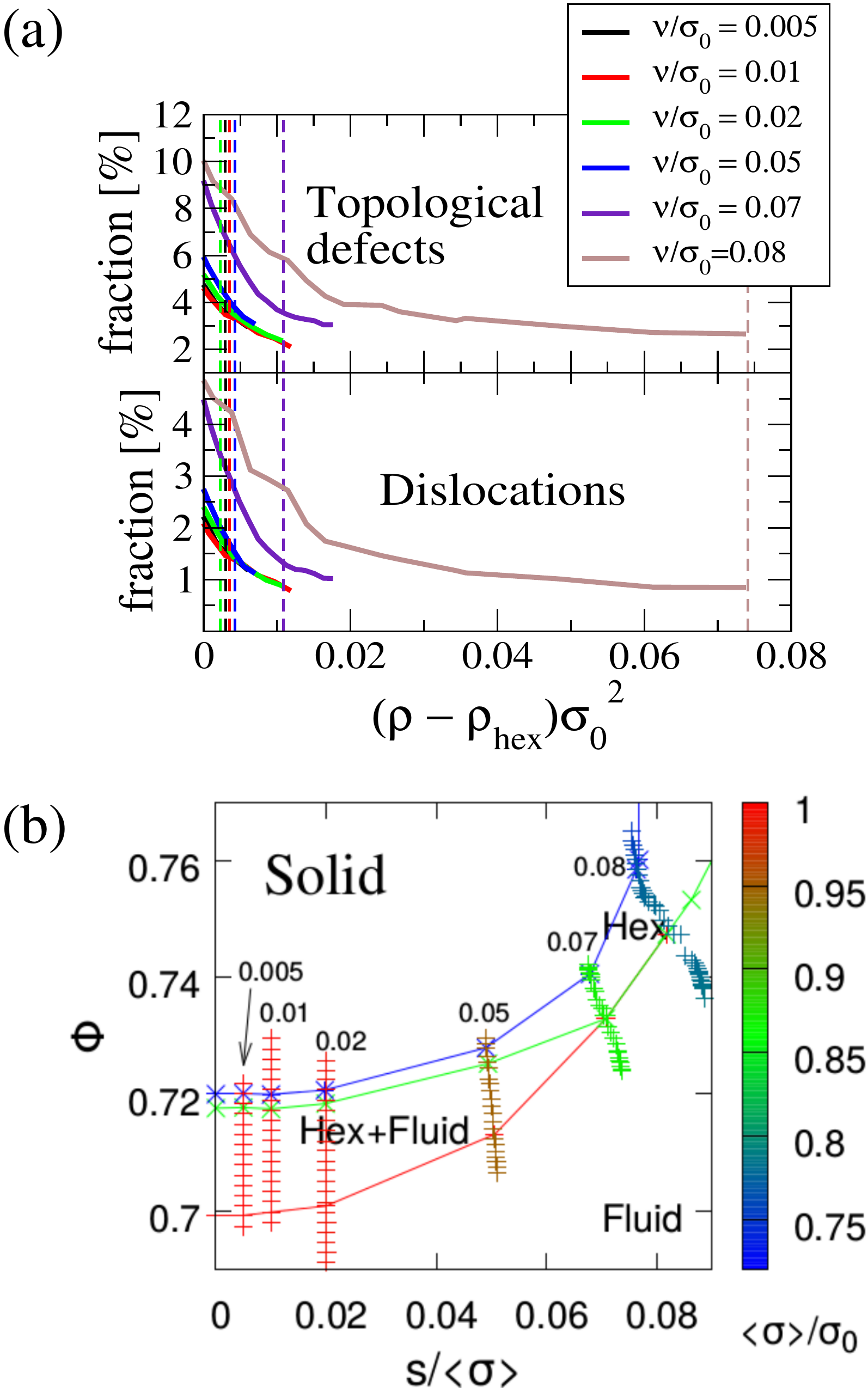}
\caption{
\textbf{Stabilization of hexatic phase.} (a): The fractions of topological defects and dislocations in the hexatic phase as functions of $(\rho - \rho_{hex})\sigma_0^2$ for systems of various polydispersity parameter $\nu$. 
Here $\rho$ and $\rho_{hex}$ are the density of the system and the lowest density of stable hexatic phase, respectively.
The vertical dashed lines indicate the hexatic-solid transition points.
(b): Low pressure phase diagram of polydisperse hard disks in the representation of $\phi$ vs $s/\langle \sigma \rangle$, where $\phi$ is the packing fraction of the system with $\sigma_i$ the diameter of particle $i$. The state points obtained from simulations at each $\nu/\sigma_0$ from 0.005 to 0.08 are shown as the symbols, which are color coded with $\langle \sigma \rangle / \sigma_0$. The error bars are smaller than the symbols. 
}
\label{fig2}
\end{figure}

\begin{figure*}
\centering \includegraphics[width=\textwidth]{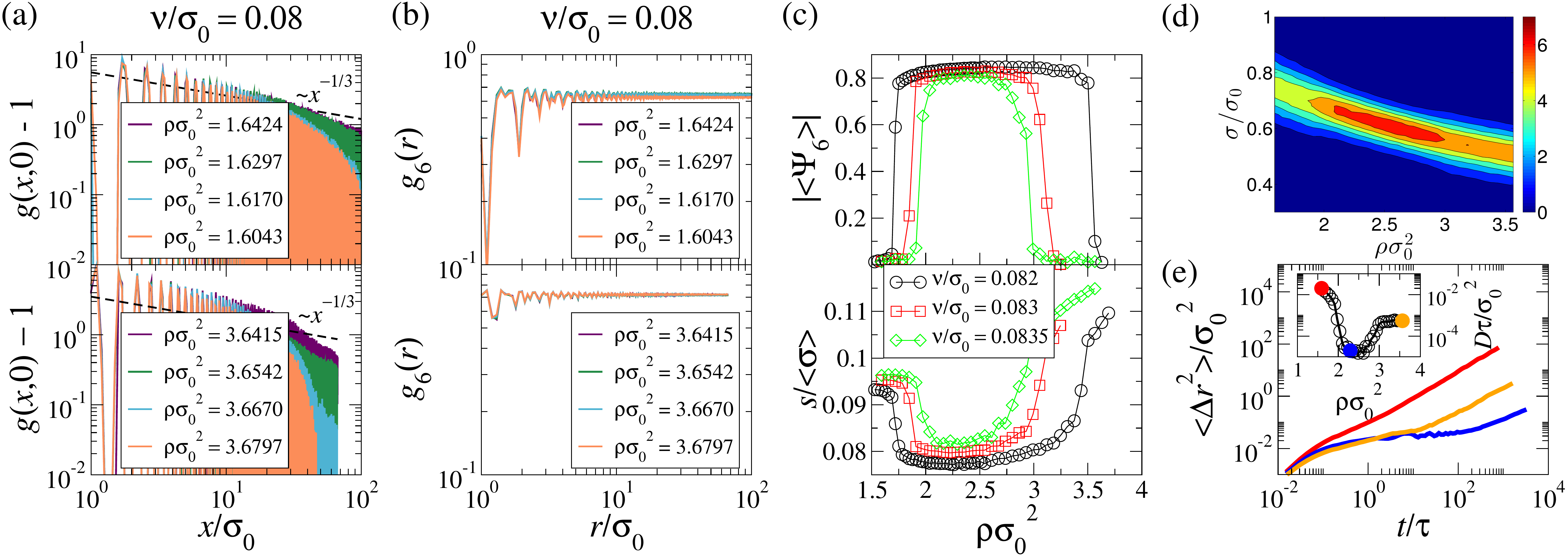}
\caption{
\textbf{Re-entrant melting transitions.} (a,b): The pair correlation function $g(x,0)-1$ along the major axis of the system (a) and the six fold orientation correlation function $g_6(r)$ (b) for polydisperse hard disk systems (PHDS) with $\nu/\sigma_0 = 0.08$ at various densities.
(c): $\left |\langle \Psi_6 \rangle \right| $ and $s/\langle \sigma \rangle$ as functions of density $\rho \sigma_0^2$ for PHDS with $\nu/\sigma_0 = 0.082$, 0.083, and 0.0835.
(d): Contour plot of the probability density distribution of particle size $\sigma/\sigma_0$ for systems of different densities at $\nu/\sigma_0 = 0.0835$.
(e): Mean square displacement $\langle \Delta r^2 \rangle / \sigma_0^2$ in PHDS of $\nu/\sigma_0 = 0.0835$ at densities $\rho \sigma_0^2 = 1.592$, 2.292, and 3.565 as marked in the inset, where $\tau = \sigma_0 \sqrt{m/k_BT}$ is the time unit of molecular dynamics simulations with $m$ the mass of particles. Inset: the diffusion coefficient $D$ as a function of density in PHDS with $\nu/\sigma_0 = 0.0835$.
}
\label{fig3}
\end{figure*}

\subsection{Phase diagram}
By performing Monte Carlo (MC) simulations, we calculate the equation of state (EOS) for a system of $256^2=65,536$ hard disks with various $\nu$ from 0.005 to $0.08\sigma_0$. As shown in Fig.~\ref{fig1}a, one can see that when the polydispersity parameter is small, i.e., $\nu/\sigma_0 \le 0.05$, there is clearly a Mayer-Wood loop in the EOS~\cite{mayerloop} implying a first-order transition from fluid with increasing the density of the system. Similar to the monodisperse hard-disk system, i.e., $\nu = 0$, the coexisting phase with fluid in PHDS is a hexatic phase~\cite{hdprl,hdpre,frenkel2004}. 
To characterize the structural difference between fluid and hexatic phases, we calculate the six-fold bond orientation order parameter $\langle \Psi_6 \rangle = \left \langle \frac{1}{N} \sum_{k=1}^N \psi_6(\mathbf{r}_k) \right \rangle$ with
$\psi_6(\mathbf{r}_k)  = \frac{1}{N_k} \sum_{j=1}^{N_k} \exp(i6 \theta_{kj})$,
where $\theta_{kj}$ is the angle between the vector connecting particle $k$ with its neighbour $j$ and a chosen fixed reference vector. $N_k$ is the number of first neighbours for particle $k$ based on the Voronoi tessellation of the system. As shown in Fig.~\ref{fig1}a, the density difference between the coexisting hexatic and fluid phases becomes smaller with increasing $\nu$, and when $\nu/\sigma_0 \ge 0.07$, the density jump from fluid to hexatic phase disappears, while $\left |\langle \Psi_6 \rangle \right|$ increases significantly at $\rho_{hex}$ (Fig.~\ref{fig1}b). 
Here we note that $\left | \langle \Psi_6 \rangle \right |$ in an infinitely large system, should be zero for hexatic phase, while in our large finite system it can be a positive number indicating the existence of some orientational order in the system. However, we do not use $\left | \langle \Psi_6 \rangle \right |$ to identify the existence of hexatic phase, for which we always check the change of correlation functions to confirm.
This suggests that the first-order fluid-hexatic transition becomes weaker with increasing $\nu$, and changes to be continuous at high polydispersity, i.e., $\nu/\sigma_0 \ge 0.07$, following the celebrated KTHNY scenario~\cite{kt,hn2,hn,young}. This, to some extent, is similar to the melting of soft spheres~\cite{krauth2015}. Because with increasing the particle size polydispersity, the distribution of distance between the nearest neighbors becomes wider, which is similar to introducing a ``soft'' repulsion between particles, and  it has been also found in the melting of soft spheres in 2D, the transition type can switch from a first order transition to a continuous KTHNY scenario with increasing the ``softness'' of the potential ~\cite{krauth2015}.

\subsection{Finite size effect on the melting in polydisperse hard disks}
It was shown that the system finite size effect is important in studying 2D melting~\cite{hdprl,schilling2011}. Therefore, we perform MC simulations for PHDS of various numbers of particles from $N=64^2$ to $512^2$ at $\nu/\sigma_0 = 0.07$, and the EOSs for different system sizes are shown in Fig.~\ref{fig1}c. When the system size is small, i.e., $N=64^2=4096$, there is a pronounced Mayer-Wood loop in EOS, signaturing a first-order fluid-hexatic transition consistent with Ref~\cite{frenkel2004}. Interestingly, with increasing the system size, the first-order fluid-hexatic transition becomes weaker, and changes to be continuous at $N \ge 256^2$. Further increasing $N$ does not change EOS significantly, which suggests the finite size effect negligible in our system of $N = 256^2$ particles. The change from a first order like transition to a continuous transition with increasing system size can be interpreted as following. A finite 2D system with periodic boundary conditions can be seen as a system wrapped on a torus, and this increases the cooperation between neighbouring particles as a result of extra ``communication'' via paths encircling the torus, which was shown being able to change the transition type from continuous to first order like in small systems~\cite{fisher1969}. By contrast, this effect does not appear in systems with open free boundaries, while in real experiments, the open boundary walls can induce order in the fluid. Moreover, for $\nu/\sigma_0 < 0.07$, we ensure that most of the coexisting densities do not change significantly with further increasing the system size to $N=512^2$, while the only exception is for systems at $\nu/\sigma_0 = 0.05$ (see Supplementary Figure 1), of which the exact boundary is out of the reach with our present computation capability.

\subsection{Stabilization of the hexatic phase}
With increasing the density of the system, the hexatic phase solidifies, and the hexatic-solid transition point can be obtained by checking the pair correlation function $g(x,0) - 1$ along the major axis of system switching from an exponential decay to a power law decay upon solidification~\cite{hdprl}.
The resulting phase diagram is plotted in the representation of $s/\langle \sigma \rangle$ vs $\rho \sigma_0^2$ in Fig.~\ref{fig1}d, where $s = \sqrt{\langle \sigma^2 \rangle - \langle \sigma \rangle^2 }$ is the actual particle size polydispersity in the system. One can see that with increasing the polydispersity of the system, the density range for stable hexatic phase increases substantially for $s/\langle \sigma \rangle$ above 0.07. 

To understand the physics behind this enhanced stability of hexatic phase in PHDS, we calculate the fraction of topological defects in the system using the method in Ref~\cite{glotzer2017}. Particle $k$ is identified as a topological defect if its disclination charge $q_a = N_k - 6 \ne 0$. These topological defects form clusters which can be classified based on the total disclination charge and Burgers vector~\cite{glotzer2017}. Defect clusters having nonzero Burgers vectors and zero disclination charges are called dislocations, and the dislocations as well as the defect clusters with nonzero disclination charges can destroy the bond orientational order in 2D solids.
Although the total amount of topological defects is not directly related to the stability of hexatic phase,
the dislocations are evidence for the hexatic phase along with  
few defect clusters with nonzero disclination charges~\cite{glotzer2017}. The fraction of particles in defect clusters with nonzero disclination charge is much smaller, i.e., more than one order of magnitude smaller, than that of dislocations in the hexatic phase found in our simulations, although the changing trends of these two types of defects are very similar (see Supplementary Figure 2).

The calculated fractions of topological defects and dislocations in our simulations with various $\nu/\sigma_0$ are plotted as functions of $(\rho - \rho_{hex})\sigma_0^2$ in Fig~\ref{fig2}a, where $\rho_{hex}$ is the lowest density of stable hexatic phase, and the vertical dashed lines locate hexatic-solid transitions. One can see that typically when the fraction of dislocations decreases to below about $1 \sim 1.5\%$, the system solidifies. When the polydispersity is small, i.e., $\nu/\sigma_0 \le 0.05$, the dependence of both fractions of defects and dislocations on density do not change significantly with increasing $\nu$, and the density ranges for stable hexatic phase below the dashed lines in Fig.~\ref{fig2}a are almost the same. However, at $\nu/\sigma_0 \ge 0.07$, the fraction of topological defects increases to around $10\%$ at $\rho_{hex}$, and also the fraction of dislocations increases to about $4.5\%$ at $\rho_{hex}$. Along with increasing the density at $\nu/\sigma_0 = 0.07$ and 0.08, the fraction of dislocations decreases to below the threshold for solidification at much higher density compared with the system of $\nu/\sigma_0 \le 0.05$. The fraction of defect clusters with nonzero disclination charges also changes similarly in systems of different polydispersity (Supplementary Figure 2).
Therefore, this suggests that the size polydispersity of hard disks creates more topological defects, which could subsequently increase the fraction of dislocations as well as the defect clusters with nonzero disclination charges in the system. This destroys the quasi-long-range positional correlation in the solid driving the formation of hexatic phase. 
As the chemical potential difference $\Delta \mu (\sigma)$ is fixed in our simulations, the average particle size in the system decreases with increasing density. Therefore, the packing fraction of the system $\phi = \frac{\pi}{4} \langle \sum_{i} \sigma_i^2/V \rangle$ does not linearly increase with the density, while the actual particle size distribution remains very close to a Gaussian-like distribution centered around $\langle\sigma\rangle$ (see Supplementary Figure 3).
At very high density, when the polydispersity is high enough, e.g., $\nu/\sigma_0 \simeq 0.08$, the packing fraction of the system can even decrease with increasing the density in $\mu VT$ ensemble, which different from the $NVT$ ensemble~\cite{santen200}. This decreasing packing fraction at high density signatures a re-entrant melting transition, which we explain later. In experiments, one of the most relevant parameters is the packing fraction of the system. Thus we plot the low pressure phase diagram of PHDS in the representation of $\phi$ vs $s/\langle \sigma \rangle$ in Fig.~\ref{fig2}b. Here the low pressure means the range of pressure close to the fluid$\rightarrow$hexatic and hexatic$\rightarrow$solid transitions below the re-entrant transitions. One can see that in systems of larger size polydispersity, the average particle size is smaller, and the actual packing fraction range for stable hexatic increases from $0.2 \sim 0.3\%$ in the monodisperse hard-disk system to about $2 \sim 3\%$ in the PHDS with $s/\langle \sigma \rangle \simeq 0.07$. 
Moreover, at fixed $\nu$, increasing the density of the system along with the fluid-hexatic and hexatic-solid transitions decreases the actual polydispersity $s/\langle \sigma \rangle$ in the system, which is due to the formation of more ordered structures eliminating the particle size fluctuation in the system.

\subsection{Re-entrant meltings}
Another interesting feature of the phase diagram in Fig.~\ref{fig1}d is that when the size polydispersity of hard disks is around 0.08 to 0.1, at very high density, increasing density triggers re-entrant melting transitions from solid to hexatic and hexatic to fluid. 
As shown in Fig.~\ref{fig3}a, for PHDS with $\nu/\sigma_0 = 0.08$, with increasing the density from $\rho\sigma_0^2 = 1.6$ to 1.6424, the system transforms from a hexatic phase with the short-range positional correlation, i.e., an exponential decay $g(x,0)-1 \sim \exp(-x)$, to a solid with quasi-long-range positional order, i.e., a power law decay $g(x,0)-1 \sim x^{-\alpha}$ with $\alpha \le 1/3$, where the six fold orientation correlation function $g_6(r) = \langle \psi_6^*(\mathbf{r}'+\mathbf{r})\psi_6(\mathbf{r}')\rangle$~\cite{weikai2014} remains almost the same (Fig.~\ref{fig3}b). At much higher density, increasing the density from $\rho\sigma_0^2 = 3.64$ to 3.67, $g(x,0)-1$ changes again from a power law decay to an exponential decay with almost the same $g_6(r)$ (Fig.~\ref{fig3}b), which suggests a re-entrant transition from solid to hexatic phase. For hard-disk systems with higher polydispersity, i.e., $\nu/\sigma_0 = 0.082 \sim 0.0835$, we plot $\left | \langle \Psi_6 \rangle \right |$ as functions of $\rho \sigma_0^2 $ in Fig.~\ref{fig3}c, and one can see that with increasing $\rho$, $\left | \langle \Psi_6 \rangle \right |$ first increases from 0 to around 0.8 indicating the formation of an ordered phase, and further increasing $\rho$ leads to the drop of $\left | \langle \Psi_6 \rangle \right |$ to 0 implying a re-entrant melting transition into a disordered phase. By checking the scaling of $g(x,0)-1$ in the system, we ensure that the ordered phase formed is a hexatic phase. 

To understand the mechanism behind the re-entrant melting of hexatic phase, we plot $s/\langle \sigma \rangle$ as functions of $\rho \sigma_0^2$ in the lower panel of Fig.~\ref{fig3}c, from which one can find a clear correlation between the change of $\left | \langle \Psi_6 \rangle \right |$ and $s/\langle \sigma \rangle$. 
Here we ensure that our simulations at high density have been well equilibrated and independent with the initial configurations (Supplementary Figure 5). Actually, it has been shown that the equilibration in systems of continuously polydisperse particles is much faster than that in monodisperse systems~\cite{ludovicprl2016,ludovicprx2017,ikeda2017}.  
At high density, during the decrease of $\left | \langle \Psi_6 \rangle \right |$, $s/\langle \sigma \rangle$ increases significantly to the value even higher than that in the low density fluid.
This suggests that by increasing the density of the system, the standard deviation of particle size increases, and the combinatorial entropy in the system associated with the variation of particle size increases, which stabilizes the amorphous structures against the ordered ones in the system.

Moreover, along with the decrease of $\left | \langle \Psi_6 \rangle \right |$ with density at high pressure, we do not observe any Mayer-Wood loops in EOS (Supplementary Figure 4). This implies that the re-entrant melting of hexatic phase at high density is continuous, which is similar to that in soft particle systems~\cite{zuprl2016,ryzhov}.
Furthermore, it was shown that the strong particle size fractionation in polydisperse hard spheres eliminates the possibility of re-entrant melting in 3D polydisperse hard-sphere solids~\cite{sollich2003prl}. Therefore, in Fig.~\ref{fig3}d, we plot the probability density distribution of particle size $p(\sigma)$ for PHDS of various density with $\nu/\sigma_0 = 0.0835$. One can see that within the whole density range, there is always a single peak in $p(\sigma)$ at fixed $\rho$, indicating no particle size fractionation in the system. This also suggests that  re-entrant melting can indeed exist in 2D polydisperse particle systems, if there is no particle size fractionation~\cite{bartlett1999}.  Here we note that although all our simulations starting with different particle size distributions converge to the same result at fixed polydispersity and density, there still exist a possibility that the system may posses an equilibrium multimodal particle size distribution which is not accessible with direct simulations~\cite{bartlett1999}.

In addition, to check whether the amorphous phase formed at high density is a fluid or glass, we perform event driven molecular dynamics (EDMD) simulations starting from the equilibrated configurations obtained from our MC simulations at $\nu/\sigma_0 = 0.0835$.
In the EDMD simulation, the system evolves via a time-ordered sequence of elastic collision events, which are
described by Newton's equations of motion. The spheres move
at constant velocities between collisions, and the velocities of the respective particles are updated when a collision occurs. All collisions are elastic and preserve energy and momentum~\cite{rapaportmd}.
The calculated mean square displacement (MSD) for various densities are shown in Fig.~\ref{fig3}e. One can see that not only the diffusion coefficient increases by nearly two orders of magnitude along with the re-entrant melting of hexatic phase (Fig.~\ref{fig3}e inset), but also the plateau in MSD almost disappears. This suggests that the high density amorphous phase is a diffusive fluid, and in the equilibrium sedimentation of such PHDS of Gaussian-like particle size distribution, there should be a ``floating hexatic phase'' sandwiched in two fluids. 
For PHDS with $\nu/\sigma_0 \ge 0.084$, with increasing the density, we do not observe any ordered phase in our simulations.

\section{Discussions}
By performing large scale computer simulations, we investigate the effect of particle size polydispersity on the phase behaviour of hard-disk systems. We find that with increasing the size polydispersity of hard disks from zero, the first-order melting transition from hexatic phase to fluid becomes weaker, and completely switches to be continuous at $\nu/\sigma_0 \ge 0.07$ following the celebrated KTHNY scenario. Simultaneously, the density range for stable hexatic phase gets substantially enlarged. Compared with monodisperse hard-disk systems, where hexatic phase is only stable in a very small range of packing fraction of $0.2 \sim 0.3\%$~\cite{hdprl}, the packing fraction range for stable hexatic phase in PHDS with $s/\langle \sigma \rangle \simeq 0.07$ is about $2 \sim 3\%$. 
This suggests new directions in searching for the hexatic phase in 2D polydisperse particle systems.
More interestingly, in PHDS with even higher polydispersity, i.e., $0.08 \le \nu/\sigma_0 \lesssim 0.0835$, we find that at very high density, increasing density of the system can trigger re-entrant transitions from solid to hexatic and hexatic to fluid phases depending on $\nu$. 
In polydisperse systems, re-entrant transitions, especially the re-entrant melting, were originally predicted theoretically for 3D hard-sphere crystals~\cite{bartlett1999}, but proven impossible due to the strong fractionation that was not taken into account in the theory~\cite{sollich2003prl}. We find that the absence of strong fractionation in 2D polydisperse systems can indeed induce re-entrant transitions at high density. These show a new difference between phase transitions in polydisperse hard-sphere systems in 2D and 3D, and suggest a new direction in investigating phase transitions in 2D systems by considering the particle size polydispersity, which was overlooked in the past.

\section{Methods}
We perform Monte Carlo simulations in the semigrand canonical ensemble ($NVT-\Delta \mu$) using a square simulation box with periodic boundary conditions in both directions, in which each particle can randomly change its diameter $\sigma$ under the control of Eq.~(\ref{eq1}). To accelerate the equilibration and sampling, we implement the rejection free event-chain MC algorithm, with which the pressure $P$ in the system can be calculated using the mean excess chain displacement~\cite{manon2013}
\begin{equation}
 P = k_B T \rho \left \langle \frac{x_{final} - x_{initial}}{L_c}\right \rangle_{chains},
\end{equation}
where $x_{final}$ and $x_{initial}$ are the initial and final positions of each event chain along the direction of the chain, respectively. $L_c$ and $\rho$ are the chain length and number density of the particles, respectively, with $\left \langle \cdot \right \rangle_{chains}$ calculating the average over all event chains. For each simulation, we perform about $10^8 - 10^9$ MC sweeps, and each MC sweep on average consists of 1 event chain with the chain length of $L_c = 100\sigma_0$ and 500 trials of randomly changing the diameter of a randomly selected particle.

\section{Author Contributions}
R.N. conceived the research. P.S.R. performed the event chain Monte Carlo simulations. Q.-L.L. performed the event driven molecular dynamics simulations. All authors analysed the results and wrote the manuscript.

\section{Competing interests}
The authors declare no competing interests.

\section{Code availability}
Codes used for the numeric simulations are available on request from R Ni at Nanyang Technological University, Singapore.

\section{Data availability}
All data that support the findings of this study are available from the corresponding author on reasonable request.

\begin{acknowledgments}
We thank Dr. Saurish Chakrabarty for helpful discussions.
This work is supported by Nanyang Technological University Start-Up Grant (NTU-SUG: M4081781.120), the Academic Research Fund from Singapore Ministry of Education (M4011616.120 and M4011873.120), and the Advanced Manufacturing and Engineering Young Individual Research Grant (A1784C0018) by the Science and Engineering Research Council of Agency for Science, Technology and Research Singapore.
We are grateful to the National Supercomputing Centre (NSCC) of Singapore for supporting the numerical calculations.
\end{acknowledgments}

%
%
%
%

\begin{thebibliography}{50}%
\makeatletter
\providecommand \@ifxundefined [1]{%
 \@ifx{#1\undefined}
}%
\providecommand \@ifnum [1]{%
 \ifnum #1\expandafter \@firstoftwo
 \else \expandafter \@secondoftwo
 \fi
}%
\providecommand \@ifx [1]{%
 \ifx #1\expandafter \@firstoftwo
 \else \expandafter \@secondoftwo
 \fi
}%
\providecommand \natexlab [1]{#1}%
\providecommand \enquote  [1]{``#1''}%
\providecommand \bibnamefont  [1]{#1}%
\providecommand \bibfnamefont [1]{#1}%
\providecommand \citenamefont [1]{#1}%
\providecommand \href@noop [0]{\@secondoftwo}%
\providecommand \href [0]{\begingroup \@sanitize@url \@href}%
\providecommand \@href[1]{\@@startlink{#1}\@@href}%
\providecommand \@@href[1]{\endgroup#1\@@endlink}%
\providecommand \@sanitize@url [0]{\catcode `\\12\catcode `\$12\catcode
  `\&12\catcode `\#12\catcode `\^12\catcode `\_12\catcode `\%12\relax}%
\providecommand \@@startlink[1]{}%
\providecommand \@@endlink[0]{}%
\providecommand \url  [0]{\begingroup\@sanitize@url \@url }%
\providecommand \@url [1]{\endgroup\@href {#1}{\urlprefix }}%
\providecommand \urlprefix  [0]{URL }%
\providecommand \Eprint [0]{\href }%
\providecommand \doibase [0]{http://dx.doi.org/}%
\providecommand \selectlanguage [0]{\@gobble}%
\providecommand \bibinfo  [0]{\@secondoftwo}%
\providecommand \bibfield  [0]{\@secondoftwo}%
\providecommand \translation [1]{[#1]}%
\providecommand \BibitemOpen [0]{}%
\providecommand \bibitemStop [0]{}%
\providecommand \bibitemNoStop [0]{.\EOS\space}%
\providecommand \EOS [0]{\spacefactor3000\relax}%
\providecommand \BibitemShut  [1]{\csname bibitem#1\endcsname}%
\let\auto@bib@innerbib\@empty
\bibitem [{\citenamefont {Strandburg}(1988)}]{Strandburg1988}%
  \BibitemOpen
  \bibfield  {author} {\bibinfo {author} {\bibfnamefont {Katherine~J.}\
  \bibnamefont {Strandburg}},\ }\bibfield  {title} {\enquote {\bibinfo {title}
  {Two-dimensional melting},}\ }\href {\doibase 10.1103/RevModPhys.60.161}
  {\bibfield  {journal} {\bibinfo  {journal} {Rev. Mod. Phys.}\ }\textbf
  {\bibinfo {volume} {60}},\ \bibinfo {pages} {161--207} (\bibinfo {year}
  {1988})}\BibitemShut {NoStop}%
\bibitem [{\citenamefont {Dash}(1999)}]{Dash1999}%
  \BibitemOpen
  \bibfield  {author} {\bibinfo {author} {\bibfnamefont {J.~G.}\ \bibnamefont
  {Dash}},\ }\bibfield  {title} {\enquote {\bibinfo {title} {History of the
  search for continuous melting},}\ }\href {\doibase
  10.1103/RevModPhys.71.1737} {\bibfield  {journal} {\bibinfo  {journal} {Rev.
  Mod. Phys.}\ }\textbf {\bibinfo {volume} {71}},\ \bibinfo {pages}
  {1737--1743} (\bibinfo {year} {1999})}\BibitemShut {NoStop}%
\bibitem [{\citenamefont {Gasser}(2009)}]{Grasser2009}%
  \BibitemOpen
  \bibfield  {author} {\bibinfo {author} {\bibfnamefont {U.}~\bibnamefont
  {Gasser}},\ }\bibfield  {title} {\enquote {\bibinfo {title} {Crystallization
  in three- and two-dimensional colloidal suspensions},}\ }\href@noop {}
  {\bibfield  {journal} {\bibinfo  {journal} {Journal of Physics: Condensed
  Matter}\ }\textbf {\bibinfo {volume} {21}},\ \bibinfo {pages} {203101}
  (\bibinfo {year} {2009})}\BibitemShut {NoStop}%
\bibitem [{\citenamefont {Kosterlitz}\ and\ \citenamefont
  {Thouless}(1973)}]{kt}%
  \BibitemOpen
  \bibfield  {author} {\bibinfo {author} {\bibfnamefont {J.M.}\ \bibnamefont
  {Kosterlitz}}\ and\ \bibinfo {author} {\bibfnamefont {D.J.}\ \bibnamefont
  {Thouless}},\ }\bibfield  {title} {\enquote {\bibinfo {title} {Ordering,
  metastability and phase transitions in two-dimensional systems},}\
  }\href@noop {} {\bibfield  {journal} {\bibinfo  {journal} {Journal of Physics
  C: Solid State Physics}\ }\textbf {\bibinfo {volume} {6}},\ \bibinfo {pages}
  {1181} (\bibinfo {year} {1973})}\BibitemShut {NoStop}%
\bibitem [{\citenamefont {Halperin}\ and\ \citenamefont {Nelson}(1978)}]{hn2}%
  \BibitemOpen
  \bibfield  {author} {\bibinfo {author} {\bibfnamefont {B.~I.}\ \bibnamefont
  {Halperin}}\ and\ \bibinfo {author} {\bibfnamefont {David~R.}\ \bibnamefont
  {Nelson}},\ }\bibfield  {title} {\enquote {\bibinfo {title} {Theory of
  two-dimensional melting},}\ }\href@noop {} {\bibfield  {journal} {\bibinfo
  {journal} {Phys. Rev. Lett.}\ }\textbf {\bibinfo {volume} {41}},\ \bibinfo
  {pages} {121--124} (\bibinfo {year} {1978})}\BibitemShut {NoStop}%
\bibitem [{\citenamefont {Nelson}\ and\ \citenamefont {Halperin}(1979)}]{hn}%
  \BibitemOpen
  \bibfield  {author} {\bibinfo {author} {\bibfnamefont {David~R.}\
  \bibnamefont {Nelson}}\ and\ \bibinfo {author} {\bibfnamefont {B.~I.}\
  \bibnamefont {Halperin}},\ }\bibfield  {title} {\enquote {\bibinfo {title}
  {Dislocation-mediated melting in two dimensions},}\ }\href@noop {} {\bibfield
   {journal} {\bibinfo  {journal} {Phys. Rev. B}\ }\textbf {\bibinfo {volume}
  {19}},\ \bibinfo {pages} {2457--2484} (\bibinfo {year} {1979})}\BibitemShut
  {NoStop}%
\bibitem [{\citenamefont {Young}(1979)}]{young}%
  \BibitemOpen
  \bibfield  {author} {\bibinfo {author} {\bibfnamefont {A.~P.}\ \bibnamefont
  {Young}},\ }\bibfield  {title} {\enquote {\bibinfo {title} {Melting and the
  vector coulomb gas in two dimensions},}\ }\href@noop {} {\bibfield  {journal}
  {\bibinfo  {journal} {Phys. Rev. B}\ }\textbf {\bibinfo {volume} {19}},\
  \bibinfo {pages} {1855--1866} (\bibinfo {year} {1979})}\BibitemShut {NoStop}%
\bibitem [{\citenamefont {Berezinskii}(1971)}]{bkt}%
  \BibitemOpen
  \bibfield  {author} {\bibinfo {author} {\bibfnamefont {V.}~\bibnamefont
  {Berezinskii}},\ }\bibfield  {title} {\enquote {\bibinfo {title} {Destruction
  of long range order in one-dimensional and two-dimensional systems having a
  continuous symmetry group. i. classical systems},}\ }\href@noop {} {\bibfield
   {journal} {\bibinfo  {journal} {Sov. Phys. JETP}\ }\textbf {\bibinfo
  {volume} {32}},\ \bibinfo {pages} {493} (\bibinfo {year} {1971})}\BibitemShut
  {NoStop}%
\bibitem [{\citenamefont {Binder}\ \emph {et~al.}(2002)\citenamefont {Binder},
  \citenamefont {Sengupta},\ and\ \citenamefont {Nielaba}}]{binder2002}%
  \BibitemOpen
  \bibfield  {author} {\bibinfo {author} {\bibfnamefont {Kurt}\ \bibnamefont
  {Binder}}, \bibinfo {author} {\bibfnamefont {Surajit}\ \bibnamefont
  {Sengupta}}, \ and\ \bibinfo {author} {\bibfnamefont {Peter}\ \bibnamefont
  {Nielaba}},\ }\bibfield  {title} {\enquote {\bibinfo {title} {The
  liquid-solid transition of hard discs: first-order transition or
  kosterlitz-thouless-halperin-nelson-young scenario?}}\ }\href@noop {}
  {\bibfield  {journal} {\bibinfo  {journal} {Journal of Physics: Condensed
  Matter}\ }\textbf {\bibinfo {volume} {14}},\ \bibinfo {pages} {2323}
  (\bibinfo {year} {2002})}\BibitemShut {NoStop}%
\bibitem [{\citenamefont {Chui}(1982)}]{chiu1982}%
  \BibitemOpen
  \bibfield  {author} {\bibinfo {author} {\bibfnamefont {S.~T.}\ \bibnamefont
  {Chui}},\ }\bibfield  {title} {\enquote {\bibinfo {title} {Grain-boundary
  theory of melting in two dimensions},}\ }\href@noop {} {\bibfield  {journal}
  {\bibinfo  {journal} {Phys. Rev. Lett.}\ }\textbf {\bibinfo {volume} {48}},\
  \bibinfo {pages} {933--935} (\bibinfo {year} {1982})}\BibitemShut {NoStop}%
\bibitem [{\citenamefont {Saito}(1982)}]{saito1982}%
  \BibitemOpen
  \bibfield  {author} {\bibinfo {author} {\bibfnamefont {Y.}~\bibnamefont
  {Saito}},\ }\bibfield  {title} {\enquote {\bibinfo {title} {Melting of
  dislocation vector systems in two dimensions},}\ }\href@noop {} {\bibfield
  {journal} {\bibinfo  {journal} {Phys. Rev. Lett.}\ }\textbf {\bibinfo
  {volume} {48}},\ \bibinfo {pages} {1114--1117} (\bibinfo {year}
  {1982})}\BibitemShut {NoStop}%
\bibitem [{\citenamefont {Zahn}\ \emph {et~al.}(1999)\citenamefont {Zahn},
  \citenamefont {Lenke},\ and\ \citenamefont {Maret}}]{zahn1999}%
  \BibitemOpen
  \bibfield  {author} {\bibinfo {author} {\bibfnamefont {K.}~\bibnamefont
  {Zahn}}, \bibinfo {author} {\bibfnamefont {R.}~\bibnamefont {Lenke}}, \ and\
  \bibinfo {author} {\bibfnamefont {G.}~\bibnamefont {Maret}},\ }\bibfield
  {title} {\enquote {\bibinfo {title} {Two-stage melting of paramagnetic
  colloidal crystals in two dimensions},}\ }\href@noop {} {\bibfield  {journal}
  {\bibinfo  {journal} {Phys. Rev. Lett.}\ }\textbf {\bibinfo {volume} {82}},\
  \bibinfo {pages} {2721--2724} (\bibinfo {year} {1999})}\BibitemShut {NoStop}%
\bibitem [{\citenamefont {Karnchanaphanurach}\ \emph
  {et~al.}(2000)\citenamefont {Karnchanaphanurach}, \citenamefont {Lin},\ and\
  \citenamefont {Rice}}]{karn2000}%
  \BibitemOpen
  \bibfield  {author} {\bibinfo {author} {\bibfnamefont {Pallop}\ \bibnamefont
  {Karnchanaphanurach}}, \bibinfo {author} {\bibfnamefont {Binhua}\
  \bibnamefont {Lin}}, \ and\ \bibinfo {author} {\bibfnamefont {Stuart~A.}\
  \bibnamefont {Rice}},\ }\bibfield  {title} {\enquote {\bibinfo {title}
  {Melting transition in a quasi-two-dimensional colloid suspension: Influence
  of the colloid-colloid interaction},}\ }\href@noop {} {\bibfield  {journal}
  {\bibinfo  {journal} {Phys. Rev. E}\ }\textbf {\bibinfo {volume} {61}},\
  \bibinfo {pages} {4036--4044} (\bibinfo {year} {2000})}\BibitemShut {NoStop}%
\bibitem [{\citenamefont {Han}\ \emph {et~al.}(2008)\citenamefont {Han},
  \citenamefont {Ha}, \citenamefont {Alsayed},\ and\ \citenamefont
  {Yodh}}]{han2008}%
  \BibitemOpen
  \bibfield  {author} {\bibinfo {author} {\bibfnamefont {Y.}~\bibnamefont
  {Han}}, \bibinfo {author} {\bibfnamefont {N.~Y.}\ \bibnamefont {Ha}},
  \bibinfo {author} {\bibfnamefont {A.~M.}\ \bibnamefont {Alsayed}}, \ and\
  \bibinfo {author} {\bibfnamefont {A.~G.}\ \bibnamefont {Yodh}},\ }\bibfield
  {title} {\enquote {\bibinfo {title} {Melting of two-dimensional
  tunable-diameter colloidal crystals},}\ }\href@noop {} {\bibfield  {journal}
  {\bibinfo  {journal} {Phys. Rev. E}\ }\textbf {\bibinfo {volume} {77}},\
  \bibinfo {pages} {041406} (\bibinfo {year} {2008})}\BibitemShut {NoStop}%
\bibitem [{\citenamefont {Rice}(2009)}]{RICE20091}%
  \BibitemOpen
  \bibfield  {author} {\bibinfo {author} {\bibfnamefont {Stuart~A.}\
  \bibnamefont {Rice}},\ }\bibfield  {title} {\enquote {\bibinfo {title}
  {Structure in confined colloid suspensions},}\ }\href@noop {} {\bibfield
  {journal} {\bibinfo  {journal} {Chemical Physics Letters}\ }\textbf {\bibinfo
  {volume} {479}},\ \bibinfo {pages} {1 -- 13} (\bibinfo {year}
  {2009})}\BibitemShut {NoStop}%
\bibitem [{\citenamefont {Murray}\ and\ \citenamefont
  {Van~Winkle}(1987)}]{murray1987}%
  \BibitemOpen
  \bibfield  {author} {\bibinfo {author} {\bibfnamefont {C.~A.}\ \bibnamefont
  {Murray}}\ and\ \bibinfo {author} {\bibfnamefont {D.~H.}\ \bibnamefont
  {Van~Winkle}},\ }\bibfield  {title} {\enquote {\bibinfo {title} {Experimental
  observation of two-stage melting in a classical two-dimensional screened
  coulomb system},}\ }\href@noop {} {\bibfield  {journal} {\bibinfo  {journal}
  {Phys. Rev. Lett.}\ }\textbf {\bibinfo {volume} {58}},\ \bibinfo {pages}
  {1200--1203} (\bibinfo {year} {1987})}\BibitemShut {NoStop}%
\bibitem [{\citenamefont {Marcus}\ and\ \citenamefont
  {Rice}(1996)}]{marcus1996}%
  \BibitemOpen
  \bibfield  {author} {\bibinfo {author} {\bibfnamefont {Andrew~H.}\
  \bibnamefont {Marcus}}\ and\ \bibinfo {author} {\bibfnamefont {Stuart~A.}\
  \bibnamefont {Rice}},\ }\bibfield  {title} {\enquote {\bibinfo {title}
  {Observations of first-order liquid-to-hexatic and hexatic-to-solid phase
  transitions in a confined colloid suspension},}\ }\href@noop {} {\bibfield
  {journal} {\bibinfo  {journal} {Phys. Rev. Lett.}\ }\textbf {\bibinfo
  {volume} {77}},\ \bibinfo {pages} {2577--2580} (\bibinfo {year}
  {1996})}\BibitemShut {NoStop}%
\bibitem [{\citenamefont {von Gr\"unberg}\ \emph {et~al.}(2004)\citenamefont
  {von Gr\"unberg}, \citenamefont {Keim}, \citenamefont {Zahn},\ and\
  \citenamefont {Maret}}]{maret2004}%
  \BibitemOpen
  \bibfield  {author} {\bibinfo {author} {\bibfnamefont {H.~H.}\ \bibnamefont
  {von Gr\"unberg}}, \bibinfo {author} {\bibfnamefont {P.}~\bibnamefont
  {Keim}}, \bibinfo {author} {\bibfnamefont {K.}~\bibnamefont {Zahn}}, \ and\
  \bibinfo {author} {\bibfnamefont {G.}~\bibnamefont {Maret}},\ }\bibfield
  {title} {\enquote {\bibinfo {title} {Elastic behavior of a two-dimensional
  crystal near melting},}\ }\href@noop {} {\bibfield  {journal} {\bibinfo
  {journal} {Phys. Rev. Lett.}\ }\textbf {\bibinfo {volume} {93}},\ \bibinfo
  {pages} {255703} (\bibinfo {year} {2004})}\BibitemShut {NoStop}%
\bibitem [{\citenamefont {Keim}\ \emph {et~al.}(2007)\citenamefont {Keim},
  \citenamefont {Maret},\ and\ \citenamefont {von Gr\"unberg}}]{keim2007}%
  \BibitemOpen
  \bibfield  {author} {\bibinfo {author} {\bibfnamefont {P.}~\bibnamefont
  {Keim}}, \bibinfo {author} {\bibfnamefont {G.}~\bibnamefont {Maret}}, \ and\
  \bibinfo {author} {\bibfnamefont {H.~H.}\ \bibnamefont {von Gr\"unberg}},\
  }\bibfield  {title} {\enquote {\bibinfo {title} {Frank's constant in the
  hexatic phase},}\ }\href@noop {} {\bibfield  {journal} {\bibinfo  {journal}
  {Phys. Rev. E}\ }\textbf {\bibinfo {volume} {75}},\ \bibinfo {pages} {031402}
  (\bibinfo {year} {2007})}\BibitemShut {NoStop}%
\bibitem [{\citenamefont {Xu}\ and\ \citenamefont {Rice}(2008)}]{stuart2008}%
  \BibitemOpen
  \bibfield  {author} {\bibinfo {author} {\bibfnamefont {Xinliang}\
  \bibnamefont {Xu}}\ and\ \bibinfo {author} {\bibfnamefont {Stuart~A.}\
  \bibnamefont {Rice}},\ }\bibfield  {title} {\enquote {\bibinfo {title}
  {Liquid-to-hexatic phase transition in a quasi-two-dimensional colloid
  system},}\ }\href@noop {} {\bibfield  {journal} {\bibinfo  {journal} {Phys.
  Rev. E}\ }\textbf {\bibinfo {volume} {78}},\ \bibinfo {pages} {011602}
  (\bibinfo {year} {2008})}\BibitemShut {NoStop}%
\bibitem [{\citenamefont {Bernard}\ and\ \citenamefont {Krauth}(2011)}]{hdprl}%
  \BibitemOpen
  \bibfield  {author} {\bibinfo {author} {\bibfnamefont {Etienne~P.}\
  \bibnamefont {Bernard}}\ and\ \bibinfo {author} {\bibfnamefont {Werner}\
  \bibnamefont {Krauth}},\ }\bibfield  {title} {\enquote {\bibinfo {title}
  {Two-step melting in two dimensions: First-order liquid-hexatic
  transition},}\ }\href {\doibase 10.1103/PhysRevLett.107.155704} {\bibfield
  {journal} {\bibinfo  {journal} {Phys. Rev. Lett.}\ }\textbf {\bibinfo
  {volume} {107}},\ \bibinfo {pages} {155704} (\bibinfo {year}
  {2011})}\BibitemShut {NoStop}%
\bibitem [{\citenamefont {Engel}\ \emph {et~al.}(2013)\citenamefont {Engel},
  \citenamefont {Anderson}, \citenamefont {Glotzer}, \citenamefont {Isobe},
  \citenamefont {Bernard},\ and\ \citenamefont {Krauth}}]{hdpre}%
  \BibitemOpen
  \bibfield  {author} {\bibinfo {author} {\bibfnamefont {Michael}\ \bibnamefont
  {Engel}}, \bibinfo {author} {\bibfnamefont {Joshua~A.}\ \bibnamefont
  {Anderson}}, \bibinfo {author} {\bibfnamefont {Sharon~C.}\ \bibnamefont
  {Glotzer}}, \bibinfo {author} {\bibfnamefont {Masaharu}\ \bibnamefont
  {Isobe}}, \bibinfo {author} {\bibfnamefont {Etienne~P.}\ \bibnamefont
  {Bernard}}, \ and\ \bibinfo {author} {\bibfnamefont {Werner}\ \bibnamefont
  {Krauth}},\ }\bibfield  {title} {\enquote {\bibinfo {title} {Hard-disk
  equation of state: First-order liquid-hexatic transition in two dimensions
  with three simulation methods},}\ }\href {\doibase
  10.1103/PhysRevE.87.042134} {\bibfield  {journal} {\bibinfo  {journal} {Phys.
  Rev. E}\ }\textbf {\bibinfo {volume} {87}},\ \bibinfo {pages} {042134}
  (\bibinfo {year} {2013})}\BibitemShut {NoStop}%
\bibitem [{\citenamefont {Kapfer}\ and\ \citenamefont
  {Krauth}(2015)}]{krauth2015}%
  \BibitemOpen
  \bibfield  {author} {\bibinfo {author} {\bibfnamefont {Sebastian~C.}\
  \bibnamefont {Kapfer}}\ and\ \bibinfo {author} {\bibfnamefont {Werner}\
  \bibnamefont {Krauth}},\ }\bibfield  {title} {\enquote {\bibinfo {title}
  {Two-dimensional melting: From liquid-hexatic coexistence to continuous
  transitions},}\ }\href@noop {} {\bibfield  {journal} {\bibinfo  {journal}
  {Phys. Rev. Lett.}\ }\textbf {\bibinfo {volume} {114}},\ \bibinfo {pages}
  {035702} (\bibinfo {year} {2015})}\BibitemShut {NoStop}%
\bibitem [{\citenamefont {Anderson}\ \emph {et~al.}(2017)\citenamefont
  {Anderson}, \citenamefont {Antonaglia}, \citenamefont {Millan}, \citenamefont
  {Engel},\ and\ \citenamefont {Glotzer}}]{glotzer2017}%
  \BibitemOpen
  \bibfield  {author} {\bibinfo {author} {\bibfnamefont {Joshua~A.}\
  \bibnamefont {Anderson}}, \bibinfo {author} {\bibfnamefont {James}\
  \bibnamefont {Antonaglia}}, \bibinfo {author} {\bibfnamefont {Jaime~A.}\
  \bibnamefont {Millan}}, \bibinfo {author} {\bibfnamefont {Michael}\
  \bibnamefont {Engel}}, \ and\ \bibinfo {author} {\bibfnamefont {Sharon~C.}\
  \bibnamefont {Glotzer}},\ }\bibfield  {title} {\enquote {\bibinfo {title}
  {Shape and symmetry determine two-dimensional melting transitions of hard
  regular polygons},}\ }\href@noop {} {\bibfield  {journal} {\bibinfo
  {journal} {Phys. Rev. X}\ }\textbf {\bibinfo {volume} {7}},\ \bibinfo {pages}
  {021001} (\bibinfo {year} {2017})}\BibitemShut {NoStop}%
\bibitem [{\citenamefont {Li}\ and\ \citenamefont
  {Ciamarra}(2018)}]{massimo2018}%
  \BibitemOpen
  \bibfield  {author} {\bibinfo {author} {\bibfnamefont {Yan-Wei}\ \bibnamefont
  {Li}}\ and\ \bibinfo {author} {\bibfnamefont {Massimo~Pica}\ \bibnamefont
  {Ciamarra}},\ }\bibfield  {title} {\enquote {\bibinfo {title} {Role of cell
  deformability in the two-dimensional melting of biological tissues},}\
  }\href@noop {} {\bibfield  {journal} {\bibinfo  {journal} {Phys. Rev.
  Materials}\ }\textbf {\bibinfo {volume} {2}},\ \bibinfo {pages} {045602}
  (\bibinfo {year} {2018})}\BibitemShut {NoStop}%
\bibitem [{\citenamefont {Deutschl\"ander}\ \emph {et~al.}(2013)\citenamefont
  {Deutschl\"ander}, \citenamefont {Horn}, \citenamefont {L\"owen},
  \citenamefont {Maret},\ and\ \citenamefont {Keim}}]{lowen2013}%
  \BibitemOpen
  \bibfield  {author} {\bibinfo {author} {\bibfnamefont {Sven}\ \bibnamefont
  {Deutschl\"ander}}, \bibinfo {author} {\bibfnamefont {Tobias}\ \bibnamefont
  {Horn}}, \bibinfo {author} {\bibfnamefont {Hartmut}\ \bibnamefont {L\"owen}},
  \bibinfo {author} {\bibfnamefont {Georg}\ \bibnamefont {Maret}}, \ and\
  \bibinfo {author} {\bibfnamefont {Peter}\ \bibnamefont {Keim}},\ }\bibfield
  {title} {\enquote {\bibinfo {title} {Two-dimensional melting under quenched
  disorder},}\ }\href@noop {} {\bibfield  {journal} {\bibinfo  {journal} {Phys.
  Rev. Lett.}\ }\textbf {\bibinfo {volume} {111}},\ \bibinfo {pages} {098301}
  (\bibinfo {year} {2013})}\BibitemShut {NoStop}%
\bibitem [{\citenamefont {Qi}\ and\ \citenamefont
  {Dijkstra}(2015)}]{weikai2015}%
  \BibitemOpen
  \bibfield  {author} {\bibinfo {author} {\bibfnamefont {Weikai}\ \bibnamefont
  {Qi}}\ and\ \bibinfo {author} {\bibfnamefont {Marjolein}\ \bibnamefont
  {Dijkstra}},\ }\bibfield  {title} {\enquote {\bibinfo {title}
  {Destabilisation of the hexatic phase in systems of hard disks by quenched
  disorder due to pinning on a lattice},}\ }\href@noop {} {\bibfield  {journal}
  {\bibinfo  {journal} {Soft Matter}\ }\textbf {\bibinfo {volume} {11}},\
  \bibinfo {pages} {2852} (\bibinfo {year} {2015})}\BibitemShut {NoStop}%
\bibitem [{\citenamefont {Russo}\ and\ \citenamefont
  {Wilding}(2017)}]{russo2017}%
  \BibitemOpen
  \bibfield  {author} {\bibinfo {author} {\bibfnamefont {John}\ \bibnamefont
  {Russo}}\ and\ \bibinfo {author} {\bibfnamefont {Nigel~B.}\ \bibnamefont
  {Wilding}},\ }\bibfield  {title} {\enquote {\bibinfo {title} {Disappearance
  of the hexatic phase in a binary mixture of hard disks},}\ }\href@noop {}
  {\bibfield  {journal} {\bibinfo  {journal} {Phys. Rev. Lett.}\ }\textbf
  {\bibinfo {volume} {119}},\ \bibinfo {pages} {115702} (\bibinfo {year}
  {2017})}\BibitemShut {NoStop}%
\bibitem [{\citenamefont {Thorneywork}\ \emph {et~al.}(2017)\citenamefont
  {Thorneywork}, \citenamefont {Abbott}, \citenamefont {Aarts},\ and\
  \citenamefont {Dullens}}]{dullens2017}%
  \BibitemOpen
  \bibfield  {author} {\bibinfo {author} {\bibfnamefont {Alice~L.}\
  \bibnamefont {Thorneywork}}, \bibinfo {author} {\bibfnamefont {Joshua~L.}\
  \bibnamefont {Abbott}}, \bibinfo {author} {\bibfnamefont {Dirk G. A.~L.}\
  \bibnamefont {Aarts}}, \ and\ \bibinfo {author} {\bibfnamefont {Roel P.~A.}\
  \bibnamefont {Dullens}},\ }\bibfield  {title} {\enquote {\bibinfo {title}
  {Two-dimensional melting of colloidal hard spheres},}\ }\href {\doibase
  10.1103/PhysRevLett.118.158001} {\bibfield  {journal} {\bibinfo  {journal}
  {Phys. Rev. Lett.}\ }\textbf {\bibinfo {volume} {118}},\ \bibinfo {pages}
  {158001} (\bibinfo {year} {2017})}\BibitemShut {NoStop}%
\bibitem [{\citenamefont {Kawasaki}\ \emph {et~al.}(2007)\citenamefont
  {Kawasaki}, \citenamefont {Araki},\ and\ \citenamefont
  {Tanaka}}]{tanaka2007}%
  \BibitemOpen
  \bibfield  {author} {\bibinfo {author} {\bibfnamefont {T.}~\bibnamefont
  {Kawasaki}}, \bibinfo {author} {\bibfnamefont {T.}~\bibnamefont {Araki}}, \
  and\ \bibinfo {author} {\bibfnamefont {H.}~\bibnamefont {Tanaka}},\
  }\bibfield  {title} {\enquote {\bibinfo {title} {Correlation between dynamic
  heterogeneity and medium-range order in two-dimensional glass-forming
  liquids},}\ }\href@noop {} {\bibfield  {journal} {\bibinfo  {journal} {Phys.
  Rev. Lett.}\ }\textbf {\bibinfo {volume} {99}},\ \bibinfo {pages} {215701}
  (\bibinfo {year} {2007})}\BibitemShut {NoStop}%
\bibitem [{\citenamefont {Kawasaki}\ and\ \citenamefont
  {Tanaka}(2011)}]{tanaka2011}%
  \BibitemOpen
  \bibfield  {author} {\bibinfo {author} {\bibfnamefont {T.}~\bibnamefont
  {Kawasaki}}\ and\ \bibinfo {author} {\bibfnamefont {H.}~\bibnamefont
  {Tanaka}},\ }\bibfield  {title} {\enquote {\bibinfo {title} {Structural
  signature of slow dynamics and dynamic heterogeneity in two-dimensional
  colloidal liquids: Glassy structural order},}\ }\href@noop {} {\bibfield
  {journal} {\bibinfo  {journal} {J Phys. Condens. Matter.}\ }\textbf {\bibinfo
  {volume} {23}},\ \bibinfo {pages} {194121} (\bibinfo {year}
  {2011})}\BibitemShut {NoStop}%
\bibitem [{\citenamefont {Russo}\ and\ \citenamefont
  {Tanaka}(2015)}]{tanaka2015}%
  \BibitemOpen
  \bibfield  {author} {\bibinfo {author} {\bibfnamefont {J.}~\bibnamefont
  {Russo}}\ and\ \bibinfo {author} {\bibfnamefont {H.}~\bibnamefont {Tanaka}},\
  }\bibfield  {title} {\enquote {\bibinfo {title} {Assessing the role of static
  length scales behind glassy dynamics in polydisperse hard disks},}\
  }\href@noop {} {\bibfield  {journal} {\bibinfo  {journal} {Proc. Natl Acad.
  Sci. USA}\ }\textbf {\bibinfo {volume} {112}},\ \bibinfo {pages} {6920}
  (\bibinfo {year} {2015})}\BibitemShut {NoStop}%
\bibitem [{\citenamefont {Fasolo}\ and\ \citenamefont
  {Sollich}(2003)}]{sollich2003prl}%
  \BibitemOpen
  \bibfield  {author} {\bibinfo {author} {\bibfnamefont {Moreno}\ \bibnamefont
  {Fasolo}}\ and\ \bibinfo {author} {\bibfnamefont {Peter}\ \bibnamefont
  {Sollich}},\ }\bibfield  {title} {\enquote {\bibinfo {title} {Equilibrium
  phase behavior of polydisperse hard spheres},}\ }\href {\doibase
  10.1103/PhysRevLett.91.068301} {\bibfield  {journal} {\bibinfo  {journal}
  {Phys. Rev. Lett.}\ }\textbf {\bibinfo {volume} {91}},\ \bibinfo {pages}
  {068301} (\bibinfo {year} {2003})}\BibitemShut {NoStop}%
\bibitem [{\citenamefont {Kofke}\ and\ \citenamefont
  {Glandt}(1988)}]{kofke1988}%
  \BibitemOpen
  \bibfield  {author} {\bibinfo {author} {\bibfnamefont {David~A.}\
  \bibnamefont {Kofke}}\ and\ \bibinfo {author} {\bibfnamefont {Eduardo~D.}\
  \bibnamefont {Glandt}},\ }\bibfield  {title} {\enquote {\bibinfo {title}
  {Monte carlo simulation of multicomponent equilibria in a semigrand canonical
  ensemble},}\ }\href@noop {} {\bibfield  {journal} {\bibinfo  {journal}
  {Molecular Physics}\ }\textbf {\bibinfo {volume} {64}},\ \bibinfo {pages}
  {1105--1131} (\bibinfo {year} {1988})}\BibitemShut {NoStop}%
\bibitem [{\citenamefont {Bolhuis}\ and\ \citenamefont
  {Kofke}(1996)}]{bolhuis1996}%
  \BibitemOpen
  \bibfield  {author} {\bibinfo {author} {\bibfnamefont {Peter~G.}\
  \bibnamefont {Bolhuis}}\ and\ \bibinfo {author} {\bibfnamefont {David~A.}\
  \bibnamefont {Kofke}},\ }\bibfield  {title} {\enquote {\bibinfo {title}
  {Monte carlo study of freezing of polydisperse hard spheres},}\ }\href
  {\doibase 10.1103/PhysRevE.54.634} {\bibfield  {journal} {\bibinfo  {journal}
  {Phys. Rev. E}\ }\textbf {\bibinfo {volume} {54}},\ \bibinfo {pages}
  {634--643} (\bibinfo {year} {1996})}\BibitemShut {NoStop}%
\bibitem [{\citenamefont {Kofke}\ and\ \citenamefont
  {Bolhuis}(1999)}]{kofke1999}%
  \BibitemOpen
  \bibfield  {author} {\bibinfo {author} {\bibfnamefont {David~A.}\
  \bibnamefont {Kofke}}\ and\ \bibinfo {author} {\bibfnamefont {Peter~G.}\
  \bibnamefont {Bolhuis}},\ }\bibfield  {title} {\enquote {\bibinfo {title}
  {Freezing of polydisperse hard spheres},}\ }\href {\doibase
  10.1103/PhysRevE.59.618} {\bibfield  {journal} {\bibinfo  {journal} {Phys.
  Rev. E}\ }\textbf {\bibinfo {volume} {59}},\ \bibinfo {pages} {618--622}
  (\bibinfo {year} {1999})}\BibitemShut {NoStop}%
\bibitem [{\citenamefont {Pronk}\ and\ \citenamefont
  {Frenkel}(2004)}]{frenkel2004}%
  \BibitemOpen
  \bibfield  {author} {\bibinfo {author} {\bibfnamefont {Sander}\ \bibnamefont
  {Pronk}}\ and\ \bibinfo {author} {\bibfnamefont {Daan}\ \bibnamefont
  {Frenkel}},\ }\bibfield  {title} {\enquote {\bibinfo {title} {Melting of
  polydisperse hard disks},}\ }\href {\doibase 10.1103/PhysRevE.69.066123}
  {\bibfield  {journal} {\bibinfo  {journal} {Phys. Rev. E}\ }\textbf {\bibinfo
  {volume} {69}},\ \bibinfo {pages} {066123} (\bibinfo {year}
  {2004})}\BibitemShut {NoStop}%
\bibitem [{\citenamefont {Mayer}\ and\ \citenamefont {Wood}(1965)}]{mayerloop}%
  \BibitemOpen
  \bibfield  {author} {\bibinfo {author} {\bibfnamefont {Joseph~E.}\
  \bibnamefont {Mayer}}\ and\ \bibinfo {author} {\bibfnamefont {Wm.~W.}\
  \bibnamefont {Wood}},\ }\bibfield  {title} {\enquote {\bibinfo {title}
  {Interfacial tension effects in finite, periodic, two‐dimensional
  systems},}\ }\href@noop {} {\bibfield  {journal} {\bibinfo  {journal} {The
  Journal of Chemical Physics}\ }\textbf {\bibinfo {volume} {42}},\ \bibinfo
  {pages} {4268--4274} (\bibinfo {year} {1965})}\BibitemShut {NoStop}%
\bibitem [{\citenamefont {Gribova}\ \emph {et~al.}(2011)\citenamefont
  {Gribova}, \citenamefont {Arnold}, \citenamefont {Schilling},\ and\
  \citenamefont {Holm}}]{schilling2011}%
  \BibitemOpen
  \bibfield  {author} {\bibinfo {author} {\bibfnamefont {Nadezhda}\
  \bibnamefont {Gribova}}, \bibinfo {author} {\bibfnamefont {Axel}\
  \bibnamefont {Arnold}}, \bibinfo {author} {\bibfnamefont {Tanja}\
  \bibnamefont {Schilling}}, \ and\ \bibinfo {author} {\bibfnamefont
  {Christian}\ \bibnamefont {Holm}},\ }\bibfield  {title} {\enquote {\bibinfo
  {title} {How close to two dimensions does a lennard-jones system need to be
  to produce a hexatic phase?}}\ }\href@noop {} {\bibfield  {journal} {\bibinfo
   {journal} {The Journal of Chemical Physics}\ }\textbf {\bibinfo {volume}
  {135}},\ \bibinfo {pages} {054514} (\bibinfo {year} {2011})}\BibitemShut
  {NoStop}%
\bibitem [{\citenamefont {Ferdinand}\ and\ \citenamefont
  {Fisher}(1969)}]{fisher1969}%
  \BibitemOpen
  \bibfield  {author} {\bibinfo {author} {\bibfnamefont {Arthur~E.}\
  \bibnamefont {Ferdinand}}\ and\ \bibinfo {author} {\bibfnamefont
  {Michael~E.}\ \bibnamefont {Fisher}},\ }\bibfield  {title} {\enquote
  {\bibinfo {title} {Bounded and inhomogeneous ising models. i. specific-heat
  anomaly of a finite lattice},}\ }\href@noop {} {\bibfield  {journal}
  {\bibinfo  {journal} {Phys. Rev.}\ }\textbf {\bibinfo {volume} {185}},\
  \bibinfo {pages} {832--846} (\bibinfo {year} {1969})}\BibitemShut {NoStop}%
\bibitem [{\citenamefont {Santen}\ and\ \citenamefont
  {Krauth}(2000)}]{santen200}%
  \BibitemOpen
  \bibfield  {author} {\bibinfo {author} {\bibfnamefont {L.}~\bibnamefont
  {Santen}}\ and\ \bibinfo {author} {\bibfnamefont {W.}~\bibnamefont
  {Krauth}},\ }\bibfield  {title} {\enquote {\bibinfo {title} {Absence of
  thermodynamic phase transition in a model glass former},}\ }\href@noop {}
  {\bibfield  {journal} {\bibinfo  {journal} {Nature}\ }\textbf {\bibinfo
  {volume} {405}},\ \bibinfo {pages} {500} (\bibinfo {year}
  {2000})}\BibitemShut {NoStop}%
\bibitem [{\citenamefont {Qi}\ \emph {et~al.}(2014)\citenamefont {Qi},
  \citenamefont {Gantapara},\ and\ \citenamefont {Dijkstra}}]{weikai2014}%
  \BibitemOpen
  \bibfield  {author} {\bibinfo {author} {\bibfnamefont {Weikai}\ \bibnamefont
  {Qi}}, \bibinfo {author} {\bibfnamefont {A.P.}\ \bibnamefont {Gantapara}}, \
  and\ \bibinfo {author} {\bibfnamefont {Marjolein}\ \bibnamefont {Dijkstra}},\
  }\bibfield  {title} {\enquote {\bibinfo {title} {Two-stage melting induced by
  dislocations and grain boundaries in monolayers of hard spheres},}\
  }\href@noop {} {\bibfield  {journal} {\bibinfo  {journal} {Soft Matter}\
  }\textbf {\bibinfo {volume} {10}},\ \bibinfo {pages} {5449} (\bibinfo {year}
  {2014})}\BibitemShut {NoStop}%
\bibitem [{\citenamefont {Berthier}\ \emph {et~al.}(2016)\citenamefont
  {Berthier}, \citenamefont {Coslovich}, \citenamefont {Ninarello},\ and\
  \citenamefont {Ozawa}}]{ludovicprl2016}%
  \BibitemOpen
  \bibfield  {author} {\bibinfo {author} {\bibfnamefont {Ludovic}\ \bibnamefont
  {Berthier}}, \bibinfo {author} {\bibfnamefont {Daniele}\ \bibnamefont
  {Coslovich}}, \bibinfo {author} {\bibfnamefont {Andrea}\ \bibnamefont
  {Ninarello}}, \ and\ \bibinfo {author} {\bibfnamefont {Misaki}\ \bibnamefont
  {Ozawa}},\ }\bibfield  {title} {\enquote {\bibinfo {title} {Equilibrium
  sampling of hard spheres up to the jamming density and beyond},}\ }\href@noop
  {} {\bibfield  {journal} {\bibinfo  {journal} {Phys. Rev. Lett.}\ }\textbf
  {\bibinfo {volume} {116}},\ \bibinfo {pages} {238002} (\bibinfo {year}
  {2016})}\BibitemShut {NoStop}%
\bibitem [{\citenamefont {Ninarello}\ \emph {et~al.}(2017)\citenamefont
  {Ninarello}, \citenamefont {Berthier},\ and\ \citenamefont
  {Coslovich}}]{ludovicprx2017}%
  \BibitemOpen
  \bibfield  {author} {\bibinfo {author} {\bibfnamefont {Andrea}\ \bibnamefont
  {Ninarello}}, \bibinfo {author} {\bibfnamefont {Ludovic}\ \bibnamefont
  {Berthier}}, \ and\ \bibinfo {author} {\bibfnamefont {Daniele}\ \bibnamefont
  {Coslovich}},\ }\bibfield  {title} {\enquote {\bibinfo {title} {Models and
  algorithms for the next generation of glass transition studies},}\
  }\href@noop {} {\bibfield  {journal} {\bibinfo  {journal} {Phys. Rev. X}\
  }\textbf {\bibinfo {volume} {7}},\ \bibinfo {pages} {021039} (\bibinfo {year}
  {2017})}\BibitemShut {NoStop}%
\bibitem [{\citenamefont {Ikeda}\ \emph {et~al.}(2017)\citenamefont {Ikeda},
  \citenamefont {Zamponi},\ and\ \citenamefont {Ikeda}}]{ikeda2017}%
  \BibitemOpen
  \bibfield  {author} {\bibinfo {author} {\bibfnamefont {Harukuni}\
  \bibnamefont {Ikeda}}, \bibinfo {author} {\bibfnamefont {Francesco}\
  \bibnamefont {Zamponi}}, \ and\ \bibinfo {author} {\bibfnamefont {Atsushi}\
  \bibnamefont {Ikeda}},\ }\bibfield  {title} {\enquote {\bibinfo {title} {Mean
  field theory of the swap monte carlo algorithm},}\ }\href@noop {} {\bibfield
  {journal} {\bibinfo  {journal} {The Journal of Chemical Physics}\ }\textbf
  {\bibinfo {volume} {147}},\ \bibinfo {pages} {234506} (\bibinfo {year}
  {2017})}\BibitemShut {NoStop}%
\bibitem [{\citenamefont {Zu}\ \emph {et~al.}(2016)\citenamefont {Zu},
  \citenamefont {Liu}, \citenamefont {Tong},\ and\ \citenamefont
  {Xu}}]{zuprl2016}%
  \BibitemOpen
  \bibfield  {author} {\bibinfo {author} {\bibfnamefont {Mengjie}\ \bibnamefont
  {Zu}}, \bibinfo {author} {\bibfnamefont {Jun}\ \bibnamefont {Liu}}, \bibinfo
  {author} {\bibfnamefont {Hua}\ \bibnamefont {Tong}}, \ and\ \bibinfo {author}
  {\bibfnamefont {Ning}\ \bibnamefont {Xu}},\ }\bibfield  {title} {\enquote
  {\bibinfo {title} {Density affects the nature of the hexatic-liquid
  transition in two-dimensional melting of soft-core systems},}\ }\href
  {\doibase 10.1103/PhysRevLett.117.085702} {\bibfield  {journal} {\bibinfo
  {journal} {Phys. Rev. Lett.}\ }\textbf {\bibinfo {volume} {117}},\ \bibinfo
  {pages} {085702} (\bibinfo {year} {2016})}\BibitemShut {NoStop}%
\bibitem [{\citenamefont {Dudalov}\ \emph {et~al.}(2014)\citenamefont
  {Dudalov}, \citenamefont {Fomin}, \citenamefont {Tsiok},\ and\ \citenamefont
  {Ryzhov}}]{ryzhov}%
  \BibitemOpen
  \bibfield  {author} {\bibinfo {author} {\bibfnamefont {D.E.}\ \bibnamefont
  {Dudalov}}, \bibinfo {author} {\bibfnamefont {Yu.D.}\ \bibnamefont {Fomin}},
  \bibinfo {author} {\bibfnamefont {E.N.}\ \bibnamefont {Tsiok}}, \ and\
  \bibinfo {author} {\bibfnamefont {V.N.}\ \bibnamefont {Ryzhov}},\ }\bibfield
  {title} {\enquote {\bibinfo {title} {Anomalous melting scenario of the
  two-dimensional core-softened system},}\ }\href@noop {} {\bibfield  {journal}
  {\bibinfo  {journal} {arXiv:1311.7534v4}\ } (\bibinfo {year}
  {2014})}\BibitemShut {NoStop}%
\bibitem [{\citenamefont {Bartlett}\ and\ \citenamefont
  {Warren}(1999)}]{bartlett1999}%
  \BibitemOpen
  \bibfield  {author} {\bibinfo {author} {\bibfnamefont {Paul}\ \bibnamefont
  {Bartlett}}\ and\ \bibinfo {author} {\bibfnamefont {Patrick~B.}\ \bibnamefont
  {Warren}},\ }\bibfield  {title} {\enquote {\bibinfo {title} {Reentrant
  melting in polydispersed hard spheres},}\ }\href {\doibase
  10.1103/PhysRevLett.82.1979} {\bibfield  {journal} {\bibinfo  {journal}
  {Phys. Rev. Lett.}\ }\textbf {\bibinfo {volume} {82}},\ \bibinfo {pages}
  {1979--1982} (\bibinfo {year} {1999})}\BibitemShut {NoStop}%
\bibitem [{\citenamefont {Rapaport}(2004)}]{rapaportmd}%
  \BibitemOpen
  \bibfield  {author} {\bibinfo {author} {\bibfnamefont {D.~C.}\ \bibnamefont
  {Rapaport}},\ }\href@noop {} {\emph {\bibinfo {title} {The Art of Molecular
  Dynamics Simulation}}}\ (\bibinfo  {publisher} {Cambridge University Press, New York},\
  \bibinfo {year} {2004})\BibitemShut {NoStop}%
\bibitem [{\citenamefont {Michel}\ \emph {et~al.}(2013)\citenamefont {Michel},
  \citenamefont {Karpfer},\ and\ \citenamefont {Krauth}}]{manon2013}%
  \BibitemOpen
  \bibfield  {author} {\bibinfo {author} {\bibfnamefont {M.}~\bibnamefont
  {Michel}}, \bibinfo {author} {\bibfnamefont {S.C.}\ \bibnamefont {Karpfer}},
  \ and\ \bibinfo {author} {\bibfnamefont {W.}~\bibnamefont {Krauth}},\
  }\bibfield  {title} {\enquote {\bibinfo {title} {Generalized event-chain
  monte carlo: Constructing rejection-free global-balance algorithms from in
  nitesimal steps},}\ }\href@noop {} {\bibfield  {journal} {\bibinfo  {journal}
  {J. Chem. Phys}\ ,\ \bibinfo {pages} {054116}} (\bibinfo {year}
  {2013})}\BibitemShut {NoStop}%
\end{thebibliography}

%

\end{document}